\documentstyle[preprint, aps]{revtex}
\widetext
\draft
\tighten
\begin{document}
\title{A generalisation of  classical electrodynamics for the prediction of scalar field effects}
\bigskip
\author{\bf Koen J. van Vlaenderen}
\address {Institute for Basic Research \\ koenvanvlaenderen@wanadoo.nl}
\date{\today}
\maketitle
\baselineskip 7mm

\bibliographystyle{amsplain}

\begin{abstract}
\noindent 
Within the framework of Classical Electrodynamics (CED) it is common practice to choose freely an arbitrary gauge condition with respect to a gauge transformation of the electromagnetic potentials. The Lorenz gauge condition allows for the derivation of the inhomogeneous potential wave equations (IPWE), but this also means that scalar derivatives of the electromagnetic potentials are considered to be \emph{unphysical}. However, these scalar expressions might have the meaning of  a new physical field, $\mathsf S$. If this is the case, then a generalised CED is required such that scalar field effects are predicted and such that experiments can be performed in order to verify or falsify this generalised CED. The IPWE are viewed as a generalised Gauss law and a generalised Ampe\`re law, that also contain derivatives of $\mathsf S$, after reformulating the IPWE in terms of fields. 

Since charge is conserved,  scalar field $\mathsf S$ satisfies the homogeneous wave equation, thus one should expect primarily sources of dynamic scalar fields, and not sources of static scalar fields. The collective tunneling of charges might be an exception to this, since quantum tunneling is the quantum equivalent of a classical local violation of charge continuity. Generalised power/force theorems are derived that are useful in order to review historical experiments since the beginning of  electrical engineering, for instance Nikola Tesla's high voltage high frequency experiments.  Longitudinal electro-scalar vacuum waves, longitudinal forces that act on current elements, and applied power by means of static charge and the $\mathsf S$ field, are predicted by this theory. The energy density and field stress terms of scalar field $\mathsf S$ are defined.

Some recent  experiment show positive results that are in qualitative agreement with the presented predictions of scalar field effects, but further quantitative tests are required in order to verify or falsify the presented theory. The importance of Nikola Tesla's pioneering research, with respect to the predicted effects, cannot be overstated.
\end{abstract}
 
\pagebreak


\section{Introduction}

\noindent
In general, the Maxwell/Heaviside equations, completed by the Lorentz force law, are considered to be a complete theory for classical electrodynamics \cite{maxwell}. In differential form these quations are:
\begin{eqnarray}
\mathsf	\vec \nabla  \!\! \cdot \!\! \vec E  \quad          & = &      \quad  \frac{\rho}{\epsilon _0}  \quad  \quad \quad   \rm Gauss~law  \\
\mathsf 	\vec \nabla  \!\! \times \!\! \vec B  \; -  \epsilon _0 \mu _0  \frac{\partial \vec E}{\partial t}   \quad          & = & 
\mathsf          		\quad  \mu _0 \vec J    \quad  \quad \quad  \rm Amp\grave ere ~law \\
\mathsf	\vec \nabla  \!\! \times \!\! \vec E \; +  \frac{\partial \vec B}{\partial t} \quad            & = &  \mathsf      \quad \vec 0  \quad \quad \quad   \rm Faraday~law  \\
\mathsf	\vec \nabla  \!\! \cdot \!\! \vec B \quad            & = &    \mathsf      \quad  0
\end{eqnarray}
The electromagnetic fields $\mathsf \vec E$ and $\mathsf \vec B$, and an extra scalar expression $\mathsf S$, can be defined in terms of the electromagnetic potentials, $\mathsf \Phi$ and $\mathsf \vec A$:

\begin{eqnarray}
\mathsf 	\vec B  \quad &=& \mathsf \quad   \vec \nabla  \!\! \times \!\! \vec A  \\
\mathsf 	\vec E \quad &=& \mathsf \quad  - \vec \nabla  \Phi  \;   - \frac{\partial \vec A}{\partial t}  \\
\mathsf 	       S \quad  &=&  \mathsf \quad  - \epsilon _0 \mu _0  \frac{\partial \Phi}{\partial t} \;  -  \vec \nabla  \!\! \cdot \!\! \vec A
\end{eqnarray}
In terms of the potentials and expressions S, the Gauss law and the Amp\`ere law are:
\begin{eqnarray}
\mathsf 	 \left( \epsilon _0 \mu _0  \frac{\partial ^2 \Phi}{\partial t ^2}  \; - \vec \nabla  ^2 \Phi \right)
	   \; +       \frac{\partial S}{\partial t}  \quad  & = &   \mathsf      \quad  \frac{\rho}{\epsilon _0}  \\
\mathsf 	 \left( \epsilon _0 \mu _0  \frac{\partial ^2 \vec A}{\partial t ^2} \; - \vec \nabla  ^2 \vec A \right)
	  \; -      \vec \nabla  S  \quad  & = &        \mathsf  \quad  \mu _0 \vec J
\end{eqnarray}
The Maxwell/Heaviside equations are \emph{invariant} with respect to a gauge transformation, defined by a scalar function  $\chi$:
\begin{eqnarray}
\mathsf 	\Phi &\longrightarrow &\mathsf \Phi ' \quad  = \quad \Phi + \frac{\partial \chi}{\partial t} \\
\mathsf 	\vec A & \longrightarrow & \mathsf \vec A ' \quad = \quad \vec A - \vec \nabla  \chi \\
\mathsf 	\vec B & \longrightarrow & \mathsf \vec B ' \quad = \quad \vec B \\
\mathsf 	\vec E & \longrightarrow & \mathsf \vec E ' \quad = \quad \vec E \\
\mathsf 	S & \longrightarrow & \mathsf S' \mathsf \quad = \quad S - \left( \epsilon _0 \mu _0  \frac{\partial ^2 \chi}{\partial t ^2}  - \vec \nabla  ^2 \chi \right) \label{eqnS'}
\end{eqnarray}
because the electromagnetic fields $\mathsf \vec E$ and $\mathsf \vec B$  are invariant with respect to this transformation, and  the Maxwell/Heaviside equations do not contain partial derivatives of S. This means that for each physical situation there is not a unique solution for the potentials $\mathsf \Phi$ and $\mathsf \vec A$, because a particular solution for $\mathsf \Phi$ and $\mathsf \vec A$ can be transformed into many other solutions via an arbitrary scalar function $\chi$. From the set of all equivalent electromagnetic potential functions,  one can choose freely a particular subset such that these potentials satisfy an extra gauge condition, such as
\begin{equation}
\mathsf 	S \quad  = \quad 0
\end{equation}
which is known as the Lorenz condition \cite{lorenz}. For potentials that satisfy $\mathsf S=0$,  equations (8) and (9) become:
\begin{eqnarray}
\mathsf 	 \epsilon _0 \mu _0  \frac{\partial ^2 \Phi}{\partial t ^2}  \; - \vec \nabla  ^2 \Phi   \quad  & = &    \mathsf     \quad  \frac{\rho}{\epsilon _0}  \label{eqnIEPWE} \\
\mathsf 	 \epsilon _0 \mu _0  \frac{\partial ^2 \vec A}{\partial t ^2} \;  - \vec \nabla  ^2 \vec A  \quad    & = &    \mathsf     \quad  \mu _0 \vec J
\end{eqnarray}
which are the inhomogeneous potential wave equations (IPWE). Well known solutions of these differential equations are the retarded potentials, and in particular the Li\'enard-Wiechert potentials \cite{lienard} \cite{wiechert}. These solutions can be further evaluated and phenomena like cyclotron radiation and synchrotron radiation can be explained by these evaluations of the IPWE. It is necessary to prove that the retarded potentials satisfy the Lorenz condition \cite{vlaenderen}, and this is this case. However, other solutions than the retarded or advanced potentials exist.

A very different philosophy is to regard the  IPWE as generalised Gauss and Amp\`ere laws. In the spirit of J.C. Maxwell, who added the famous displacement term to the Amp\`ere law, one can add derivatives of expression $S$ to the Maxwell/Heaviside equations:
\begin{eqnarray}
\mathsf 	\vec \nabla  \!\! \cdot \!\! \vec E  \;  -  \frac{\partial S}{\partial t}   \quad     & = &    \mathsf     \quad  \frac{\rho}{\epsilon _0}   \label{eqGAUSS}\\
\mathsf 	\vec \nabla  \!\! \times \!\! \vec B  \;  + \vec \nabla  S \;   -  \epsilon _0 \mu _0  \frac{\partial \vec E}{\partial t}   \quad          & = & \mathsf   \quad  \mu _0 \vec J     \label{eqAMPERE} \\
\mathsf 	\vec \nabla  \!\! \times \!\! \vec E \; + \frac{\partial \vec B}{\partial t} \quad            & = & \mathsf         \quad \vec 0  \\
\mathsf 	\vec \nabla  \!\! \cdot \!\! \vec B  \quad       & = &   \mathsf     \quad  0
\end{eqnarray}
When these extra derivatives of $\mathsf S$ are likewise added to equations (8) and (9), this yield automatically the IPWE without the need of an extra gauge condition. These field equations are a generalisation of classical electrodynamics, since the special case $\mathsf S = 0$  results into the usual  Maxwell/Heaviside equations, and they are variant with respect to an arbitrary scalar gauge transformation $\mathsf \chi$, see Eq.~(\ref{eqnS'}), unless $\mathsf \chi$ is a solution of the homogeneous wave equation. The expression $\mathsf S$ now has the meaning of a physical and observable scalar field. This scalar field interacts with the vector fields $\mathsf \vec E$ and $\mathsf \vec B$, as described by the generalised field equations. The question: "Is classical electrodynamics a \emph{complete} classical field theory, with respect to scalar expression $\mathsf S$?", can not be answered within the context of the standard classical electrodynamics, 
since this theory treats  $\mathsf S$ as a non-observable non-physical function, and this is premature. The usual gauge freedom and gauge condition $\mathsf S=0$ are based on the presumption that partial derivatives of S are not part of the standard Maxwell field equations in the first place, which implies that $\mathsf S$ is disregarded as a physical field even before the theoretical development of the gauge transformation. In other words, the assumed gauge freedom and free choice of gauge  conditions are part of a sequence of \emph{circular} arguments, that seem to "prove" that $\mathsf S$ has no physical relevance. Oliver Heaviside did not like the abstract electromagnetic potentials and he preferred the concept of observable fields. Therefore it is also in the spirit of Heaviside to assume that $\mathsf S$ can cause observable field effects as required by a testable theory, and such that the Lorenz condition has only meaning as a special physical condition similar to: 'the electric field is zero'.

Next, the induction of scalar fields is discussed, followed by the derivation of generalised force/power theorems in order to predict the type of observable phenomena attributable to the presence of scalar fields.


\section{The induction of scalar fields} 

\noindent Considering the definition of $\mathsf S \; (S = - \epsilon _0 \mu _0  \frac{\partial \Phi}{\partial t}  \;  -  \vec \nabla  \!\! \cdot \! \vec A)$, one might design an electrical device such that factor ${\raise0.7ex\hbox{$\mathsf {\partial \Phi }$} \!\mathord{\left/ {\vphantom {{\partial \Phi } {\partial t}}}\right.\kern-\nulldelimiterspace}\!\lower0.7ex\hbox{$\mathsf {\partial t}$}} $ or  factor $\mathsf \vec \nabla   \!\! \cdot \!  \vec A$ is optimised, and such that these two scalar factors do not cancel each other. With ${\raise0.7ex\hbox{$\mathsf {\partial \Phi }$} \!\mathord{\left/ {\vphantom {{\partial \Phi } {\partial t}}}\right.\kern-\nulldelimiterspace}\!\lower0.7ex\hbox{$\mathsf {\partial t}$}} $    we can associate systems of high voltage and high frequency, such as \emph{pulsed power} systems. With $\mathsf \vec \nabla  \!\! \cdot \!  \vec A$  we can associate a source of divergent/convergent currents, which is similar to the induction of a magnetic field by rotating currents, $\mathsf \vec B = \vec \nabla  \!\! \times \!\! \vec A$. For instance, a spherical or cylindrical capacitor can show currents with non-zero divergence/convergence. If the capacity is high, then we can expect a high $\mathsf \; \vec \nabla  \!\! \cdot \!  \vec A$, since strong currents need to charge/discharge the capacitor. If the capacity is low, then a higher factor ${\raise0.7ex\hbox{$\mathsf {\partial \Phi }$} \!\mathord{\left/ {\vphantom {{\partial \Phi } {\partial t}}}\right.\kern-\nulldelimiterspace}\!\lower0.7ex\hbox{$\mathsf {\partial t}$}} $ can be expected, since then it takes less time to charge and discharge the capacitor to high voltages.

Electromagnetic fields are of static or dynamic type. Considering the inhomogeneous field wave equations:
\begin{eqnarray}
\mathsf 	\epsilon _0 \mu _0  \frac{\partial ^2 \vec E}{\partial t^2} - \vec \nabla  ^2 \vec E \quad
\mathsf 	   & = & \mathsf  \quad  \mu_0 \left( \!\! - \frac{\partial \vec J}{\partial t} -  \frac{\vec \nabla  \rho}{\epsilon _0 \mu _0} \right)  \\
\mathsf 	\epsilon _0 \mu _0  \frac{\partial ^2 \vec B}{\partial t^2} - \vec \nabla  ^2 \vec B \quad   \mathsf 	   & = & \mathsf  \quad \mu_0 ( \vec \nabla  \!\! \times \!\! \vec J ) \\
\mathsf 	\epsilon _0 \mu _0  \frac{\partial ^2 S}{\partial t^2} - \vec \nabla  ^2 S \quad   
\mathsf 	   & = & \mathsf  \quad \mu_0 \left( \!\! -\vec \nabla  \!\! \cdot \! \vec J - \frac{\partial \rho}{\partial t} \right)   \label{SWAVE}
\end{eqnarray}
 that are deduced from the generalised Maxwell/Heaviside field equations, we can expect primarily \emph{dynamic} scalar fields, because of the conservation of charge. This is the reason why the discovery of scalar field $S$ is not as easy as the discovery of the electromagnetic fields via simple static field type experiments. Quantum tunneling of electrons can be understood on the classical level as a local violation of charge conservation, for instance at Josephson junctions. Hence, collective quantum tunneling devices might induce a new type of classical field: a static scalar field. A dynamic scalar field is induced by a charge/current density wave: set $\mathsf \vec E = \vec 0$ and $\mathsf \vec B = \vec 0$, then Eq.(\ref{eqGAUSS}) and Eq.(\ref{eqAMPERE}) become:
 \begin{eqnarray}
 \mathsf 	 -  \frac{\partial S}{\partial t}   \quad     & = &    \mathsf     \quad  \frac{\rho}{\epsilon _0}  \\
\mathsf 	\vec \nabla  S \quad         & = & \mathsf   \quad  \mu _0 \vec J 
\end{eqnarray}
Since $\mathsf S$ satisfies wave Eq.(\ref{SWAVE}), also the charge density $\mathsf \rho$ and current density $\mathsf \vec J$ are wave solutions, however these wave solutions also have speed c. There are also wave solutions of charge/current density with speed less than c, in case the electric field (and/or the magnetic field) and scalar field are not zero. Conclusion, a scalar field $\mathsf S$ can be induced by a dynamic charge/current distribution.


\section{Generalised power/force laws}

 \noindent First, a source transformation is defined in order to generalise the standard electrodynamic force and power theorems:
\begin{eqnarray}
\mathsf 	\rho & \longrightarrow  &\mathsf  \rho '               = \rho + \epsilon _0 \frac{\partial S}{\partial t} \\
\mathsf 	\vec J  & \longrightarrow  &  \mathsf  \vec J '    = \vec J  -  \frac{1}{\mu _0} \vec \nabla  S
\end{eqnarray}
This source transformation transforms the Maxwell equations into the generalised Maxwell equations. The electrodynamic power theorem and force theorem are given by:
\begin{eqnarray}
\mathsf 	\mu _0 \! \left( \vec J \!  \cdot \!  \vec E \right)    \quad  & = &  \mathsf \quad -\frac{\partial \left( \epsilon _0 \mu _0 E^2 + B^2 \right)}{2 \: \partial t} 
						\; - \vec \nabla  \! \! \cdot \!\! \left( \vec E \! \!  \times \! \!  \vec B \right)   \label{eqnPTO}    \\
\mathsf 	\mu _0 \!  \left( \rho \vec E + \vec J \!\! \times \!\! \vec B \right)    \quad  & = & \mathsf \quad  \epsilon _0 \mu _0 \! \left( (  \vec \nabla  \!\! \cdot \!\! \vec E )\vec E
						\; + (\vec \nabla  \!\! \times \!\! \vec E ) \!\! \times \!\! \vec E \right)
						 \; + (\vec \nabla  \!\! \times \!\! \vec B ) \!\! \times \!\! \vec B
						\; - \epsilon _0 \mu _0 \frac{\partial  ( \vec E \!\! \times \!\! \vec B ) }{\partial t}  \label{eqnFTO}
\end{eqnarray}
Next, the left hand side of these theorems is  transformed:
\begin{eqnarray}
\mathsf 	\mu _0 \! \left( \vec J \!  \cdot \!  \vec E \right)  \quad  &  \longrightarrow  &
\mathsf 	          \mu _0 \left(  \vec J \; - \frac{1}{\mu _0} \vec \nabla  S  \right) \!  \cdot \!  \vec E \; = \\ \nonumber
 	& & \mathsf \quad \mu _0 \! \left( \vec J \!  \cdot \!  \vec E \right)  \; - (\vec \nabla  S) \! \cdot \!\! \vec E     \; = \\ \nonumber
 	& & \mathsf \quad \mu _0 \! \left( \vec J \!  \cdot \!  \vec E \right)  \; - \vec \nabla  \!\! \cdot \! (\vec E S) \; + S \vec \nabla  \!\! \cdot \!\! \vec E     \; = \\  \nonumber
 	& & \mathsf \quad \mu _0 \! \left( \vec J \!  \cdot \!  \vec E \right)  \; - \vec \nabla  \!\! \cdot \! (\vec E S) \; 
		+ S \left( \frac{\rho}{\epsilon _0} + \frac{\partial S}{\partial t} \right)     \; = \\ \nonumber
 	& & \mathsf \quad \mu _0 \vec J \!  \cdot \!  \vec E \; + \frac{\rho}{\epsilon _0}S   \; - \vec \nabla  \!\! \cdot \! (\vec E S) \; + \frac{\partial (S^2)}{2\partial t}
\end{eqnarray}
\begin{eqnarray}
\mathsf 	\mu _0 \!  \left( \rho \vec E + \vec J \!\! \times \!\! \vec B \right)  \;    & \longrightarrow & 
\mathsf 	 \mu _0 \!  \left( (\rho + \epsilon _0 \frac{\partial S}{\partial t}) \vec E + (\vec J  -  \frac{1}{\mu _0} \vec \nabla  S) \!\! \times \!\! \vec B \right) \; =   \\ \nonumber
	& & \mathsf \mu _0 \!  \left( \rho \vec E + \vec J \!\! \times \!\! \vec B \right)
	    \; +  \epsilon_0 \mu_0 \frac{\partial S}{\partial t} \vec E \; - (\vec \nabla  S) \!\! \times \!\! \vec B            \; =  \\ \nonumber
	& & \mathsf \mu _0 \!  \left( \rho \vec E + \vec J \!\! \times \!\! \vec B \right) 
	    \;  +  \epsilon_0 \mu_0 \frac{\partial (S \vec E)}{\partial t}  \; - \vec \nabla  \!\! \times \!\! (S \vec B) 
	    \;  + S \!\! \left( - \epsilon_0 \mu_0 \frac{\partial \vec E}{\partial t} + \vec \nabla  \!\! \times\!\! \vec B \right)       \; = \\ \nonumber
	& & \mathsf \mu _0 \!  \left( \rho \vec E + \vec J \!\! \times \!\! \vec B \right) 
	    \;  +  \epsilon_0 \mu_0 \frac{\partial (S \vec E)}{\partial t}  \; - \vec \nabla  \!\! \times \!\! (S \vec B)    \;  + S ( \mu_0 \vec J -  \vec \nabla  S)      \; = \\ \nonumber
	& & \mathsf \mu _0 \!  \left( \rho \vec E + \vec J \!\! \times \!\! \vec B + \vec J S \right) 
	    \;  +  \epsilon_0 \mu_0 \frac{\partial (S \vec E)}{\partial t}  \; - \vec \nabla  \!\! \times \!\! (S \vec B)    \;  - S \vec \nabla  S
\end{eqnarray}
The  power theorem and force theorem are transformed into:
\begin{eqnarray}
\mathsf 	\mu _0  \vec J \!  \cdot \!  \vec E  + \frac{\rho}{\epsilon _0}S     \quad & = &  \mathsf  \quad
                      - \frac{\partial \left( \epsilon _0 \mu _0 E^2 + B^2 + S^2 \right)}{2 \: \partial t}  \;  -  \vec \nabla  \! \! \cdot \!\! \left( \vec E \! \!  \times \! \!  \vec B - \vec E S \right)
		\label{eqnPTN} \\
\mathsf 	\mu _0 \!  \left( \rho \vec E + \vec J \!\! \times \!\! \vec B + \vec J S \right)    \quad  & = & \mathsf  \quad
 		  \epsilon _0 \mu _0 \! \left( (  \vec \nabla  \!\! \cdot \!\! \vec E )\vec E \; + (\vec \nabla  \!\! \times \!\! \vec E ) \!\! \times \!\! \vec E \right) \\ \nonumber
 	& &   \mathsf  \quad \quad + \; (\vec \nabla  \!\! \times \!\! \vec B + \vec \nabla  S)S \; + \; (\vec \nabla  \!\! \times \!\! \vec B + \vec \nabla  S ) \!\! \times \!\! \vec B \\ \nonumber
 	& &    \mathsf \quad \quad - \;  \epsilon _0 \mu _0 \frac{\partial  ( \vec E \!\! \times \!\! \vec B + \vec E S) }{\partial t}   \label{eqnFTN}
\end{eqnarray}
The new terms in these theorems need to be interpreted.
The generalised Poynting vector is: $\mathsf \vec P = \vec E \!\! \times \!\! \vec B - \vec E S$. The power flow vector $\mathsf \vec E S$  belongs to a new type of vacuum wave, and by setting $\mathsf \vec B = \vec 0$ we can deduce the following wave equations from the generalised Maxwell/Heaviside equations:
\begin{eqnarray}
\mathsf 	\epsilon _0 \mu _0 \frac{\partial ^2 S}{\partial t^2} \;  - \vec \nabla  \!\! \cdot \!\! \vec \nabla  S  \quad & = & \mathsf \quad 0 \\
\mathsf 	\epsilon _0 \mu _0 \frac{\partial ^2 \vec E}{\partial t^2} \;  - \vec \nabla  \vec \nabla  \!\! \cdot \!\! \vec E  \mathsf \quad & = & \quad 0
\end{eqnarray}
The solution of these wave equations is a \emph{longitudinal electro-scalar wave}, or LES wave. The term $\mathsf S^2$ represents the energy density of scalar field $\mathsf S$. The interesting term $\mathsf \frac{\rho}{\epsilon_0}S$ can be interpreted as the \emph{applied power  by means of static charge $\mathsf \rho$ and a dynamic scalar field $\mathsf S$}. The new force term $\mathsf \vec J S$  is a \emph{longitudinal force} that acts on a current element $\mathsf \vec J$.  Also new \emph{magneto-scalar stress terms} appeared in the force theorem. The scalar field is like a scalar form of magnetism: it acts on current elements and it interacts with the electric field in vacuum. The derivation of these theorems was already published in \cite{vlaenderen}, by means of the biquaternion calculus \cite{hamilton}.


\section{Experimental evidence}

\subsection{Longitudinal vacuum waves}

\noindent Nikola Tesla was one of the first scientist who mentioned the existence of longitudinal electric vacuum waves. Initially he did not believe that the wireless signals discovered by Hertz were the transversal electromagnetic (TEM) waves as predicted by Maxwell. Later Tesla acknowledged TEM waves, but he also insisted on the existence of energy efficient longitudinal electric waves, applicable for the wireless transport of energy and wireless communication. Longitudinal vacuum waves were (and still are) not accepted by the physics community as a physical reality, because this type of  wave vacuum wave is not predicted by the standard theory of electrodynamics. This should be reconsidered. Tesla's patents describe wireless energy systems \cite{tesla1}  based on Tesla's resonant transformer \cite{tesla2}  and ball-shaped antennas. Tesla optimised \cite{tesla3}  the voltage and frequency of the signal in the secondary circuit of his resonant transformer by using a secondary pancake coil \cite{tesla4} with low self-induction and a secondary spherical capacitor with low capacity. The secondary circuit voltage was about a million volt or higher. In order to prevent discharges from the secondary capacitor and secondary coil, Tesla placed the spherical capacitor in a vacuum tube, and he electrically isolated the secondary coil by submerging the coil in an oil reservoir. The capacitor could be made smaller with reduced capacity, because of the reduced risk of discharge, which further enabled Tesla to apply higher frequencies and higher voltages. Obviously Tesla optimised  scalar factor ${\raise0.7ex\hbox{$\mathsf {\partial \Phi }$} \!\mathord{\left/ {\vphantom {{\partial \Phi } {\partial t}}}\right.\kern-\nulldelimiterspace}\!\lower0.7ex\hbox{${\mathsf \partial t}$}} $ , and not scalar factor $\mathsf \vec \nabla \! \cdot \! \vec A$.

Ignatiev's experiment of longitudinal electric wave transmission, by means of a large spherical antenna, confirmed the existence of longitudinal electric vacuum waves without magnetic component \cite{ignatiev}. Ignatiev discovered that the transmitted energy was unusually high. In order to explain the result of longitudinal electric wave transmission Ignatiev concluded that a modification of the Gauss law is necessary. A possible modification of Gauss' law is presented in this paper, see Eq.~(\ref{eqGAUSS}). According to Ignatiev the measured propagation speed was about 1.2c,  in fact faster than light. Factor 1.2, is still subject of debate, and the error margin in the measurement data produced by Ignatiev is reviewed. Ignatiev excluded the existence of  the magnetic field component in the transmitted wave, and this is enough reason to refer to his experiment.

Recently, Wesley and Monstein  published a paper \cite{wesley} on the wireless transmission of longitudinal electric waves, also by means of a spherical antenna, and again the authors confirmed the existence of such a wave. According to Wesley and Monstein the transmitted energy flux is proportional to:
\begin{equation}
\mathsf 	\vec P \quad =  \quad -\vec \nabla  \Phi \frac{\partial \Phi}{\partial t}  \label{eqnW1}
\end{equation}
and the field energy density is:
\begin{equation}
\mathsf 	D \quad = \quad \frac{1}{2}(\vec \nabla  \Phi)^2 + \frac{1}{2}\left( \frac{\partial \Phi}{c \partial t} \right)^2  \label{eqnW2}
\end{equation}
 which is in agreement with the defined power flow  $\mathsf \vec E S$  and the energy density of the electric and scalar fields, $\mathsf \frac{1}{2}E^2$ and $\mathsf \frac{1}{2}S^2$  (except for a factor $\epsilon_0 \mu_0$), in case we ignore the magnetic potential $\vec A$. Wesley and Monstein determined  the polarisation of the received signal, which was indeed longitudinal. However, they did not present a background theory for the presented laws  for energy flux and energy density. Wesley and Monstein claim  Eq.~(\ref{eqnW1}) and Eq.~(\ref{eqnW2})  can be derived from Eq.~(\ref{eqnIEPWE}). This is not true. Only after the introduction of a physical scalar field ${\raise0.7ex\hbox{${\mathsf  \partial \Phi }$} \!\mathord{\left/ {\vphantom {{\partial \Phi } {\partial t}}}\right.\kern-\nulldelimiterspace}\!\lower0.7ex\hbox{$\mathsf  {c \partial t}$}} $ and after the deduction of the power theorem Eq.~(\ref{eqnPTN}), is it possible to deduce Eq.~(\ref{eqnW1}) and Eq.~(\ref{eqnW2}). Since the generalised power theorem (\ref{eqnPTN}) was published already in year 2001 \cite{vlaenderen}, it is fair to assume Wesley and Monstein reduced power theorem (\ref{eqnPTN}) to the restricted form of Eq.~(\ref{eqnW1}) and Eq.~(\ref{eqnW2}), without any reference to \cite{vlaenderen}.
 
 In \cite{chubykalo} a so called Coulomb wave is described by Tzontchev, Chubykalo and Rivera-Ju\'arez: a longitudinal electric wave. According to their measurements the Coulomb interaction is not instantaneous, but it has a finite speed which is approximately c. A Coulomb potential can be decomposed into an integral sum of electric potential waves \cite{whittaker} that all have speed c. The gradient of one such an electric potential wave is a longitudinal electric wave. The integral sum of all longitudinal waves constitutes the Coulomb electric field. As a consequence, a variation in Coulomb potential spreads with velocity c, for instance during a discharge. Since the differential with respect to time of one such an electric potential wave is a scalar field wave, there is a possibility of a hidden energy flow connected with the charge in the form of longitudinal electro-scalar waves.

\subsection{Longitudinal electrodynamic forces}

\noindent Longitudinal electrodynamic forces have been observed in several experiments, for example exploding wire experiments by Jan Nasilowski \cite{nasilowski} and Peter Graneau \cite{Graneau}. According to Graneau, the pressure due to longitudinal forces would be substantially greater than the pinch pressure. Assis and Bueno\cite{assis} showed that Amp\`ere's force law cannot be discriminated from Grassmann's force law for current elements, for any closed current circuit. They conclude both laws do not describe longitudinal forces, therefore a new theory is necessary in order to explain such forces. The standard field stress tensor does not describe longitudinal forces either, see Eq.~(\ref{eqnFTO}), because a longitudinal force term at the right hand site would not be balanced by a longitudinal force term at the left hand site, and that would render the force theorem false. Longitudinal forces can be explained by the presence of a scalar field and by the generalised force theorem (\ref{eqnFTN}), expressed by the term $\mathsf \vec J S$. This force is always parallel to the direction of current density $\mathsf \vec J$. Then one should verify that a scalar field is involved, that is induced by a source of high frequency high voltage, or by  divergent/convergent currents in a conductor or plasma. Periodic longitudinal forces give rise to charge density waves and stress \cite{moyssides} \cite{peoglos} waves, and vice verse:  a non-zero current divergence is the source of a scalar field that acts longitudinally on nearby current elements, such that another area of non-zero current divergence is created, etc...  Setting $\mathsf \vec E = \vec B = \vec 0$  in the generalised Maxwell equations, leads to charge density and current density waves; in this case  $\mathsf -\frac{\partial S}{\partial t} = \frac{\rho}{\epsilon_0}$ and $\mathsf \vec \nabla  S = \mu_0 \vec J$, and since $\mathsf S$  is a wave solution, also $\mathsf \rho$ and $\mathsf \vec J$ are waves. The fraction pattern of an exploded wire is very similar to a wave, perhaps as the direct consequence of a charge/current density wave and the breaking of the metal  bond between metal atoms in  areas with very low or very high electron density. Also Amp\`ere's \cite{ampere} hairpin experiment shows areas with divergent and convergent currents: at the exact location where currents enter and leave the hairpin \cite{hillas}.

\subsection{Applied power from static charge and a scalar field}

\noindent Usually power theorem (\ref{eqnPTO}) describes that an applied power source with density  $\mathsf \vec E \! \cdot  \! \vec J$ is converted into a radiated energy flow with density $\mathsf \vec \nabla   \! \cdot \! ( \vec E \! \times \! \vec B)$ and the change in field energy  $\mathsf \frac{1}{2}E^2$  and $\mathsf \frac{1}{2}B^2$. The expression "applied power"  should be " converted power", because a conversion of electric power is not necessarily applied power. According to the generalised power theorem (31), a scalar field $\mathsf S$ can turn an object with charge density $\mathsf \rho$ into an electrical power source, and this is expressed by term $\mathsf \frac{\rho}{\epsilon_0}S$. This static charge power source is a remarkable prediction by the theory. One should look for power conversion that involves static charge rather than dynamic currents, for instance charged objects that radiate LES waves or that show changing electric field energy. Although rumours exist that this actually has been achieved, the author is not aware of any published scientific experiments with respect to this effect.


\section{Conclusions}

\noindent The introduction of gauge conditions in CED implies that scalar derivatives of the electromagnetic potentials are non-physical. This negative hypothesis cannot be tested, and it should be reversed into the testable and positive hypothesis of measurable scalar field effects, such as longitudinal electric vacuum waves, longitudinal electrodynamic forces, and energy conversions by means of static charge and a scalar field. If these effects cannot be detected in general, then finally a \emph{physical}  justification for gauge conditions has been obtained. However, there are indications that positive results have been achieved. Further quantitative tests are needed in order to obtain scientific proof for the existence of a physical scalar field $\mathsf S$, as defined in this paper. A positive quantitative verification will have enormous consequences for the science of physics. The qualifications  "unphysical" scalar photons and "unphysical" longitudinal photons are incorrect, since these qualifications require experimental proof and the usual arguments that seem to prove them are circular. This neglect of Galileo Galilei's philosophy of physics by the physics community, with respect to gauge conditions, had serious consequences for one of the most brilliant minds in history, Nikola Tesla. There are urgent reasons to review Tesla's scientific legacy, such as the need for new forms of energy and new energy technologies.

\bibliography{references}

\providecommand{\bysame}{\leavevmode\hbox to3em{\hrulefill}\thinspace}
\providecommand{\MR}{\relax\ifhmode\unskip\space\fi MR }
\providecommand{\MRhref}[2]{%
  \href{http://www.ams.org/mathscinet-getitem?mr=#1}{#2}
}
\providecommand{\href}[2]{#2}
\begin{thebibliography}{10}

\bibitem{assis}
A.K.T. Assis and M.~Bueno, \emph{Bootstrap effect in classical
  electrodynamics}, Revista Facultad de Ingenieria \textbf{7} (2000).

\bibitem{ampere}
C.~Blondel, \emph{Amp{\`e}re et la creation de l'electrodynamique},
  Bibliotheque Nationale (1982).

\bibitem{Graneau}
P.~Graneau, \emph{Amp{\`e}re-neumann electrodynamics of metals}, Hadronic Press
  (1985).

\bibitem{hamilton}
W.R. Hamilton, \emph{On a new species of imaginary quantities connected with a
  theory of quaternions}, Proceedings of the Royal Irish Academy \textbf{2}
  (1843), 424--434.

\bibitem{hillas}
A.M. Hillas, \emph{Electromagnetic jet-propulsion: non-lorentzian forces on
  currents?}, Nature \textbf{302} (1983), 271.

\bibitem{ignatiev}
G.F. Ignatiev and V.A. Leus, \emph{On a superluminal transmission at the phase
  velocities}, Instantaneous Action at a Distance, Modern Physics:Pro and
  Contra (1999), 203.

\bibitem{lienard}
A.~Li{\'e}nard, \emph{Champ {\'e}lectrique et magn{\'e}tique produit par une
  charge {\'e}lectrique}, L' {\'E}clairage {\'E}lectique \textbf{16} (1898),
  5--14, 53--59, 106--112.

\bibitem{lorenz}
L.V. Lorenz, \emph{{\"U}ber die intensit{\"a}t der schwingungen des lichts mit
  den elektrischen str{\"o}men}, Annalen der Physik und Chemie \textbf{131}
  (1867), 243--263.

\bibitem{maxwell}
J.C. Maxwell, \emph{A dynamic theory of electromagnetic field}, Royal Society
  Transactions \textbf{155} (1865), 459--512.

\bibitem{wesley}
C.~Monstein and J.P. Wesley, \emph{Observation of scalar longitudinal
  electrodynamic waves}, Europhysics Letters \textbf{59} (2002), no.~4,
  514--520.

\bibitem{moyssides}
Paul~G. Moyssides, \emph{Experimental verification of the biot-savart-lorentz
  and the ampere force laws in a closed circuit, revisited}, I.E.E.E Trans.
  Magn. \textbf{25} (1989), p.4298--4306, p.4307--4312, p.4313--4321 (3
  Articles).

\bibitem{nasilowski}
J.~Nasilovski, \emph{Unduloids and striated disintegration of wires}, vol. 111,
  Plenum, New York, 1995.

\bibitem{peoglos}
V.~Peoglos, \emph{Measurement of the magnetostatic force of a current circuit
  on a part of itself}, J. Phys. D (1988), p.1055--1061.

\bibitem{chubykalo}
Andrew E.~Chubykalo Rumen I.~Tzontchev and Juan~M. Rivera-Ju{\'a}rez,
  \emph{Coulomb interaction does not spread instantaneously}, Hadronic Journal
  \textbf{23} (2000), 401--424.

\bibitem{tesla4}
N.~Tesla, \emph{Coil for electro-magnets}, United States Patent 512,340,
  Januari 9 1894.

\bibitem{tesla2}
\bysame, \emph{Apparatus for producing electric currents of high frequency and
  potential}, United States Patent 568,176, September 22 1896.

\bibitem{tesla3}
\bysame, \emph{Means for increasing the intensity of electrical oscillations},
  United States Patent 685,012, October 22 1901.

\bibitem{tesla1}
\bysame, \emph{Method of intensifying and utilising effects transmitted through
  natural media}, United States Patent 685,953, November 5 1905.

\bibitem{vlaenderen}
K.J. van Vlaenderen and A.~Waser, \emph{Generalisation of classical
  electrodynamics to admit a scalar field and longitudinal waves}, Hadronic
  Journal \textbf{24} (2001), 609--628.

\bibitem{whittaker}
E.T. Whittaker, \emph{On the partial differential equations of mathematical
  physics}, Mathematische Annalen \textbf{57} (1903), 333--355.

\bibitem{wiechert}
E.~Wiechert, \emph{Elektrodynamische elementargesetze}, Archives
  N{\'e}erlandaises Ser.2 \textbf{4} (1900), 549--557.

\end{thebibliography}

\end{document}